\documentclass[letterpaper, 10 pt, conference]{ieeeconf}  

\usepackage[T1]{fontenc}
\usepackage{textcomp}
\usepackage{microtype}
\usepackage{booktabs} 
\usepackage{hyperref}

\usepackage{xcolor}
\usepackage{fancybox}
\usepackage{graphicx}
\usepackage{amsmath}
\usepackage{amsfonts}
\usepackage{amssymb}
\usepackage{gensymb}
\usepackage{mathtools}
\usepackage{eurosym, booktabs}
\usepackage{pgfgantt}
\usepackage{colortbl}
\usepackage{caption}
\usepackage{subcaption}
\usepackage{float}
\usepackage{algorithmicx}
\usepackage{algorithm}
\usepackage{algpseudocode}
\usepackage{siunitx}
\usepackage{bbold}
\usepackage{nicefrac}      
\usepackage{etoolbox}
\usepackage{tabularx}
\usepackage{multirow}
\usepackage[short,nocomma]{optidef}
\newcommand{\X}{\mathcal{X}}
\newcommand{\Y}{\mathcal{Y}}
\newcommand{\I}{\mathcal{I}}
\usepackage{amsmath}
\usepackage{amssymb}
\usepackage{mathtools}
\usepackage{bbm}

\IEEEoverridecommandlockouts                              

\title{\LARGE \bf
Safe Physics-Informed Machine Learning for Dynamics and Control
}

\author{J\'an Drgo\v na$^{1}$, Truong X. Nghiem$^{2}$,
and Thomas Beckers$^{3}$, Mahyar Fazlyab$^{1}$, 
Enrique Mallada$^{1}$, Colin Jones$^{4}$, \\   Draguna Vrabie$^{5}$, Steven L. Brunton$^{6}$, and Rolf Findeisen$^{7}$
\thanks{$^{1}$Johns Hopkins University, \texttt{jdrgona1@jh.edu}}
\thanks{$^{2}$University of Central Florida,   \texttt{truong.nghiem@ucf.edu}}
\thanks{$^{3}$Vanderbilt University, \texttt{thomas.beckers@vanderbilt.edu}}
\thanks{$^{4}$École Polytechnique Fédérale de Lausanne, \texttt{colin.jones@epfl.ch}}
\thanks{$^{5}$Pacific Northwest National Laboratory, \texttt{draguna.vrabie@pnnl.gov}}
\thanks{$^{6}$University of Washington, \texttt{sbrunton@uw.edu}}
\thanks{$^{7}$TU Darmstadt, \texttt{rolf.findeisen@iat.tu-darmstadt.de}}
}

\begin{document}

\maketitle
\thispagestyle{empty}
\pagestyle{empty}

\begin{abstract}
This tutorial paper focuses on safe physics-informed machine learning in the context of dynamics and control, providing a comprehensive overview of how to integrate physical models and safety guarantees. As machine learning techniques enhance the modeling and control of complex dynamical systems, ensuring safety and stability remains a critical challenge, especially in safety-critical applications like autonomous vehicles, robotics, medical decision-making, and energy systems. We explore various approaches for embedding and ensuring safety constraints, including structural priors, Lyapunov and Control Barrier Functions, predictive control, projections, and robust optimization techniques. Additionally, we delve into methods for uncertainty quantification and safety verification, including reachability analysis and neural network verification tools, which help validate that control policies remain within safe operating bounds even in uncertain environments. The paper includes illustrative examples demonstrating the implementation aspects of safe learning frameworks that combine the strengths of data-driven approaches with the rigor of physical principles, offering a path toward the safe control of complex dynamical systems. 

\end{abstract}

\section{Introduction}

The integration of physics-informed machine learning (PIML)~\cite{Truong2023,PIML2021,Thiyagalingam2021ScientificML} with safe control strategies~\cite{Dawson2023,Brunke2022,Wang2025,Wabersich2023} has gained significant attention as a means to model and govern dynamical systems while ensuring stability, robustness, and constraint satisfaction. Traditional machine learning models excel at capturing patterns from data but often lack guarantees regarding physical consistency, generalizability, and safety. This limitation is particularly critical in domains such as robotics, autonomous systems, aerospace, and energy management, where violations of physical laws or unsafe control actions can lead to catastrophic consequences. By incorporating physical principles, safety constraints, and control-theoretic foundations into machine learning models, safe PIML provides a framework for ensuring that learned representations adhere to fundamental laws of physics while maintaining reliable and interpretable control policies.  

This tutorial explores methodologies for developing safe physics-informed machine-learning models for applications in dynamics and control. The discussion includes Section II on safe learning for dynamical systems with stability guarantees,  constraint satisfaction, and uncertainty quantification; Section III on safe learning for control via learning-based model predictive control (MPC), control barrier, and Lyapunov function-based safety guarantees; and Section IV on safety verification methods via reachability analysis, constrained optimization methods, and sampling-based approaches.
Through theoretical insights and practical examples, we demonstrate how safe PIML can be applied to real-world systems, bridging the gap between data-driven learning and physics-based modeling while ensuring compliance with safety-critical requirements.

\section{Safe Learning for Dynamics}

Safe learning for dynamics refers to the process of acquiring accurate models of system behavior while maintaining constraints that prevent unstable or physically unrealistic behavior. Unlike traditional learning-based approaches that prioritize performance over safety, safe learning integrates control-theoretic principles, uncertainty quantification, and safety constraints to guide the model design. 
 There are several choices to learn differential equation models, including dynamic mode decomposition (DMD) and Koopman operator~\cite{DMD2014,Budisic2012,KORDA2018149,KoopmanMezic2020,Brunton2022,Brunton2022book}, neural ordinary differential equations (NODEs)~\cite{NODEs2018,NEURIPS2024_7ce9df1d,NEURIPS2020_2e255d2d,Djeumou2022}, explicit neural-time steppers~\cite{Legaard2023,Yuying2022}, neural state space models~\cite{GEDON2021481,Masti2018}, sparse identification of nonlinear dynamics (SINDy)~\cite{SINDY2016}. 
  In this section, we explore key methodologies for the safe and uncertainty-aware learning of dynamical systems, providing stability and constraint satisfaction guarantees.

\subsection{Stability Guarantees}
It is important to be able to promote or enforce stability properties in learned models.  This is a strong reason to employ learning techniques rooted in a state-space framework, as opposed to relying solely on deep learning.  
Promoting stability in learned models is crucial and favors state-space-based learning techniques over purely black-box deep learning. DMD fits a best-fit linear operator to time-series data, while SINDy identifies the sparsest nonlinear model from a library of candidate functions. Both use linear regression: DMD for high-dimensional state data under low-rank assumptions, and SINDy as an over-determined regression problem with sparsity-promoting regularization for interpretability. 

There are several approaches to incorporate physics and stability into these models.  
One of the unexpected benefits of the linear regression framework is that it is relatively straightforward to incorporate additional constraints on the regression variables, in this case, the model coefficients.  Very often, physics manifests itself as conserved quantities, which result in symmetries in model coefficients.  This was first recognized in the context of regression-based models by Jean-Christophe Loiseau, who made a simple modification of the SINDy algorithm to incorporate energy conservation as a constraint on the quadratic terms of the model~\cite{Loiseau2017jfm}.  Specifically, the coefficients of the quadratic terms must be skew-symmetric for energy conservation in an incompressible fluid.  Since then, several other constraints and symmetries have been incorporated.  Perhaps the most relevant for this discussion is the innovation by Kaptanoglu et al.~\cite{kaptanoglu2021promoting} that adds additional regularization terms and constrained optimization to enforce the global stability of learned models.  This is based on the global boundedness theorem of Schlegel and Noack~\cite{Schlegel2015jfm} and holds for general quadratic systems.  More general stability guarantees are the focus of ongoing work. 

In DMD, there are similar opportunities to enforce known physics and promote stable models.  The optimized DMD~\cite{Askham2018siads} uses variable projection to enforce stability in the linear models, often dramatically improving performance.  Physics-informed DMD, as explored by Baddoo et al.~\cite{baddoo2023physics}, examines several approaches for promoting and enforcing known physics, with several canonical examples of symmetries and other physical constraints on the learned linear operator.  
Recent advances in Koopman operator theory have enabled data-driven analysis of nonlinear dynamical systems through spectral decomposition~\cite{Arbabi2017,KORDA2020599}, enabling the extraction of global stability information from data~\cite{Mauroy2016}. 
Together, these approaches offer principled tools for inferring nonlinear stability and invariant structures from data. While DMD regression can sometimes be modified to enforce manifold constraints via a Procrustes problem, in practice, regularization is often used to promote proximity to the desired physical manifold.

 \paragraph{Stability-Constrained Deep Koopman Operator (DeepKO)}
The DeepKO~\cite{Lusch2018,Folkestad2020,Yeung2019,NIPS2017_3a835d32} is a popular data-driven approach that leverages deep learning to approximate the coordinate transformations in the Koopman operator, enabling the analysis and control of nonlinear dynamical systems in a linear framework.
This tutorial demonstrates learning DeepKO with stability guarantees for system identification of nonlinear dynamical systems.
The autonomous DeepKO has been recently extended to control-oriented DeepKO models by~\cite{Shi2022,Han2020,Kaiser2017DatadrivenDO}.

The general idea is to encode system observables  $y_{k}$ via a neural network $\phi_{\theta_1}$ to compute latent states $x_{k}$ in the linear coordinates. This coordinate transformation now allows the application of the linear Koopman operator $K$ to obtain latent states at the next time step $x_{k+1}$. After the rollout over the given prediction horizon $N$, the generated latent trajectories $X = \{x_1, ..., x_N \}$ are projected back to the observable space via the decoder neural network $\phi^{-1}_{\theta_3}$.
Compactly, we can represent the DeepKO model as
\begin{subequations}
    \begin{align}
    \text{Encoder:} &  && {x}_{k} = \phi_{\theta_1}(y_k) \\
    \text{Koopman Operator:} & && {x}_{k+1} = K(x_k)  \\
    \text{Decoder:} & && \hat{y}_{k+1} = \phi^{-1}_{\theta_3}(x_{k+1}) 
    \end{align}
\end{subequations}

We assume the dataset of sampled trajectories of the system dynamics $X = [\hat{x}^i_0, ..., \hat{x}^i_{N}], \, \, i \in [1, ..., m]$
where $N$ represents the prediction horizon, $m$ represents number of measured trajectories, and $i$ represents an index of the sampled trajectory.
Given this dataset, the DeepKO model can be trained as a nonlinear system identification problem
\begin{subequations}
\label{eq:koopman_loss}
    \begin{align}
    &\underset{\theta}{\text{minimize}}     && \sum_{i=1}^m \Big( \ell_{y} +\ell_{\text{lin}} + \ell_{\text{recon}}\Big) \\
    &\text{subject to}    && \hat{y}^i_{k+1} = \phi^{-1}_{\theta_3}(K^k_{\theta_2}(\phi_{\theta_1}(y_1^i))), 
    \end{align}
\end{subequations}
with individual loss function terms defined as
\begin{subequations}
    \begin{align}
 & \ell_y = \sum_{k=1}^{N} Q_y||y^i_{k+1} - \hat{y}^i_{k+1}||_2^2, \\
   &   \ell_{\text{lin}} = \sum_{k=1}^{N}  Q_x||\phi_{\theta_1}(y_{k+1}^i) - K^k\phi_{\theta_1}(y_1^i)||_2^2, \\
   &     \ell_{\text{recon}} = Q_{\text{recon}}||y^i_1 - \phi^{-1}_{\theta_3}(\phi_{\theta_1}(y_1^i))||_2^2,
    \end{align}
\end{subequations}
where $\ell_y$ defines the output trajectory prediction loss, $\ell_{\text{lin}}$ defines the latent trajectory prediction loss, and $ \ell_{\text{recon}}$ is the encoded-decoder reconstruction loss.

However, this vanilla DeepKO approach~\eqref{eq:koopman_loss} does not guarantee the stability of the learned dynamics.
One can apply different techniques~\cite{han2022desko,Mamakoukas2023,bevanda2022} to guarantee the stability of the learned Koopman operator $K$. Since $K$ is linear, the general idea for a discrete-time system is to guarantee that its eigenvalues satisfy $|\lambda_i(K)| < 1, \ \forall i$, which can be satisfied via a range of linear algebra factorizations~\cite{Drgona2022,Skomski2021,NEURIPS2021_c9dd73f5}.
One example is to factorize  $K$ via Singular Value Decomposition (SVD) as $K = U \Sigma V$, with constrained eigenvalues implemented as
\begin{equation}
\Sigma =  \text{diag}(\lambda_{\text{max}} - (\lambda_{\text{max}} - \lambda_{\text{min}}) \cdot \sigma(\boldsymbol{ \lambda})),  
\end{equation}
where $\sigma$ is a logistic sigmoid activation function, $\cdot$ is a dot product, $\boldsymbol{ \lambda}$ is a vector of eigenvalues of the linear operator, while $\lambda_{\text{max}}$ and $\lambda_{\text{min}}$ are constraints on the maximum and minimum value of SVD factorized linear operator.
In order for the SVD factorization to hold, the left and right matrices $U$ and $V$, respectively, need to be orthogonal. This can be achieved either via Householder reflectors~\cite{zhang2018stabilizinggradientsdeepneural}, or by extending the loss function~\eqref{eq:koopman_loss} via  additional penalty term
\begin{subequations}
    \begin{align}
     &  \ell_{\text{stable}} = \ell_{U} + \ell_{V}, \\
 &  \ell_{U} = || I - UU^T||_2 + || I - U^TU||_2,   \\
 &  \ell_{V} = || I - VV^T||_2 + || I - V^TV||_2.  
    \end{align} 
\end{subequations}

The open-source code demonstrating this tutorial example on a specific dynamical system can be found in the NeuroMANCER library~\cite{Neuromancer2023}.

\paragraph{Structure- and Stability-Preserved Learning-based Port-Hamiltonian Systems (PHS)}
PHS are a structured, energy-based framework for modeling and controlling physical systems, preserving fundamental conservation laws.
They are widely used in fields like robotics \cite{duongPortHamiltonianNeuralODE2024}, power systems \cite{fiazPortHamiltonianApproachPower2013}, and fluid dynamics \cite{rathSymplecticGaussianProcess2021} due to their capacity to naturally integrate interconnections and dissipative effects.

The advent of data-driven techniques and machine learning has led to the development of Learning-Based Port-Hamiltonian Systems (LB-PHS), which leverage data to learn system representations that maintain the core physical principles of PHS, such as passivity and energy conservation.
These methods integrate machine learning tools like neural networks \cite{nearyCompositionalLearningDynamical2023a}, Gaussian processes \cite{rathSymplecticGaussianProcess2021}, and physics-informed learning to provide accurate, interpretable, and generalizable models, enhancing model identification and control synthesis.
Recently, model order reduction (MOR) techniques have been developed for LB-PHS \cite{rettbergDatadrivenIdentificationLatent2024,moreschiniDatadrivenModelReduction2024} to improve the computational efficiency, especially in real-time control applications.
This section discusses learning-based approaches to modeling PHS with a focus on structure and stability preservation.

Consider a typical nonlinear PHS expressed as
\begin{equation}
\textit{PHS:}
\begin{cases}
    \dot{x} = (J(x)-R(x)) 
    \frac{\partial H}{\partial x}(x) + G(x) u, &
    \\
    y = G^\top(x)    
    \frac{\partial H}{\partial x}(x) &
\end{cases}
\label{pHS}
\end{equation}
Here, $x$ is the state vector, $u$ the control input, and $y$ the output; $H(x): \mathbb{R}^n \rightarrow \mathbb{R}_+$ denotes the Hamiltonian representing internal energy; $J(x)$ is the skew-symmetric structure matrix, $R(x)$ the symmetric dissipative matrix, and $G(x)$ the port matrix.
The dissipation property ensures that energy loss does not exceed input energy \cite{schaftPortHamiltonianSystemsTheory2014}.
Simply speaking, the interconnection of the elements in the PHS is defined by $J$, whereas the Hamiltonian $H$ characterizes their dynamical behavior. The port variables~$u$ and $y$ are conjugate variables in the sense that their duality product defines the power flows exchanged with the environment of the system, for instance, currents and voltages in electrical circuits or forces and velocities in mechanical systems.

The Port-Hamiltonian Neural Network (PHNN) represents a PHS using neural networks for the system elements
\begin{equation}
\textit{PHNN:}
\begin{cases}
    \dot{\hat{x}} = (\hat{J}(\hat x) - \hat{R}(\hat x)) 
    \frac{\partial \hat{H}}{\partial \hat x}(\hat x) + \hat{G}(\hat x) u, &
    \\
    \hat y = \hat{G}^\top(\hat x)   
    \frac{\partial \hat{H}}{\partial \hat x}(\hat x) &
\end{cases}
\label{NN_pHS}
\end{equation}
where $\hat{H}$, $\hat{J}$, $\hat{R}$, and $\hat{G}$ are approximated by neural networks and learned from data.
To ensure their properties, $\hat{J}$ and $\hat{R}$ are parameterized as $\hat{J}(\hat x) = \hat{S}(\hat x) - \hat{S}^\top(\hat x)$ and $\hat{R} = \hat{L}(\hat x) \hat{L}^\top(\hat x)$.
The PHNN thus preserves the overall structure of the PHS.
Automatic differentiation tools \cite{paszke2017automatic,baydinAutomaticDifferentiationMachine2018} are used to calculate $\frac{\partial H_n}{\partial \hat x}$.
Training the networks involves minimizing a loss function that captures the discrepancy between predicted and actual future states, which requires a differentiable ODE solver or a numerical approximation method.
An alternative approach is to obtain the derivatives $\dot x_k$ from $x_k$, for instance, by a finite difference method \cite{gengDataDrivenReducedOrderModels2025}, then minimize the loss function between \eqref{NN_pHS} and $\dot x_k$.

While PHNNs enforce passivity, they do not inherently stabilize equilibrium points. 
For example, if the PHS \eqref{pHS} admits an (asymptotically) stable equilibrium point at the origin, it implies that $\frac{\partial H}{\partial x} \big|_{x = 0} = 0$; however, this constraint is not enforced in the training of the PHNN \eqref{NN_pHS}.
For known equilibria, their respective constraints can be included in the loss function for training the PHNN.
Recent methods from \cite{sanchez-escalonillaRobustNeuralIDAPBC2024,rothStablePortHamiltonianNeural2025} propose using Input-Convex Neural Networks (ICNN) \cite{ICNN2017} to construct Hamiltonians that stabilize certain points, though the convexity assumption can be restrictive and limit solutions to local equilibria.

\subsection{Constraints Satisfaction}

In learning dynamical systems, constraint satisfaction plays a crucial role in ensuring that learned models adhere to physical, safety, and structural properties. Unlike purely data-driven approaches, which may produce unrealistic or unstable predictions, incorporating constraints helps maintain the interpretability and adherence to the underlying physical laws by the learned dynamics.
The integration of physics-based priors, such as conservation laws, symmetries, or boundary conditions, acts as an implicit regularization mechanism, guiding the learning process toward physically consistent solutions. Additionally, constraints can be explicitly enforced using hard constraints (e.g., exact conservation laws) or soft constraints (e.g., penalty terms in loss functions), ensuring that the learned correction term remains within reasonable bounds. This approach is particularly useful in scientific machine learning applications where unconstrained deep learning models may violate fundamental physical principles, leading to unrealistic or unstable predictions.

\paragraph{Structured Learning with Universal Differential Equations (UDEs)}
UDEs introduced in~\cite{Rackauckas2020} provide a natural framework for incorporating prior knowledge and constraints into neural ODEs, ensuring that learned representations remain physically meaningful and generalizable. By embedding known governing equations into the learning process, UDEs enable constraint satisfaction through structured differential equations, thereby reducing the risk of overfitting and enhancing model interpretability. 

Structure-preserving UDEs and neural DAEs~\cite{Koch2023,koch2024NDAEs,pal2025,lueg2025,neary2025,NEURIPS2022_32cc6132} provide a robust framework for modeling networked dynamical systems while ensuring that constraints such as conservation laws, stability, and sparsity are met. Many real-world systems, including biological networks, power grids, and ecological systems, exhibit structured dependencies between subsystems. Traditional machine learning models often fail to preserve such structures, leading to physically inconsistent predictions. UDEs address this challenge by integrating data-driven corrections into governing equations while enforcing known structural constraints.

Consider a predator-prey ecosystem governed by a generalized Lotka-Volterra model, where species interact over a network. While the classical equations capture growth and interactions, many interspecies effects in complex ecosystems are unknown. Such systems can be broadly described as networked dynamical systems
\begin{equation}
\dot{x}_i = f_i(x_i) + \sum_{j \neq i}^{m} A_{ij} g_{ij}(x_i, x_j),
\end{equation}
with species dynamics defined by their states $x_i$ and state derivatives $\dot{x}_i$,
where \( f_i(x_i) \) represents the self-dynamics of each species, \( g_{ij}(x_i, x_j)  \) captures interactions between connected species, \( A  \) is the adjacency matrix defining connection with neighboring species influencing $i$-th species, and $m$ is the number of species. When the precise form of \( g_{ij} \) is unknown, a UDE-based approach models these missing interactions using a neural network component \( g_{\text{NN},ij} \), leading to 
\begin{equation}
\dot{x}_i = f_i(x_i) +  \sum_{j \neq i}^{m} A_{ij} g_{\text{NN},ij}(x_i, x_j).
\end{equation}

Moreover, in the case of the unknown or time-varying structure of the interactions, the adjacency matrix $A$ can also be learned from the data alongside $g_{\text{NN},ij}$.
To ensure structure preservation, constraints such as sparsity, conservation laws, and stability conditions can incorporated. A sparsity constraint enforces realistic network connectivity by imposing \( L1 \)-regularization on \( g_{\text{NN},ij} \), ensuring that only a subset of interactions is learned. Conservation laws ensure that total biomass or energy remains bounded in the system, enforcing the constraint $ \sum_{i} \dot{x}_i = 0$. 
Additionally, stability constraints prevent nonphysical population explosions. A Lyapunov function, defined as  
$ V(x) = \sum_{i} x_i^2$,
can be used to enforce stability by ensuring that its derivative satisfies \( dV/dt \leq 0 \), thereby constraining the learned dynamics to remain physically realistic. 

This structure-preserving UDEs and neural DAEs apply to diverse domains, including neuroscience, power grids, and process control. They enable inference of missing dynamics while enforcing domain-specific constraints. By embedding physical principles into learning, they yield interpretable and generalizable models for high-dimensional and partially known systems.
The open-source code demonstrations of learning structured UDEs from time series data can be found in the popular Julia library DifferentialEquations.jl~\cite{Rackauckas2017}.

\paragraph{Physics-constrained Motion Prediction (PCMP)}
Motion prediction methods in autonomous vehicles must not only account for the agent's intent and the physical feasibility of its motion (such as its dynamics, acceleration, and steering limits) but also provide a quantified measure of uncertainty to guide risk-aware decision-making.
Physics-driven approaches and pattern/data-driven approaches have been developed for motion prediction \cite{karle_scenario_2022}.
Data-driven approaches such as long short-term memory (LSTM) networks, graph neural networks, and transformers predict future trajectories based on historical observations and environmental context \cite{alahi_social_2016,gilles_thomas_2022,liang_learning_2020,nayakanti_wayformer_2022}.
However, these models are often vulnerable to noise, distribution shifts, and adversarial perturbations, which compromise safety.
On the other hand, physics-driven approaches use knowledge of the agent's physical dynamics to make predictions \cite{karle_scenario_2022}.
These models, such as the Constant Turn Rate and Velocity (CTRV) model and reachability-based models \cite{pek_using_2020,koschi_spot_2017}, ensure physical feasibility but suffer from limited robustness to noisy or complex contexts.
A major challenge in motion prediction is to ensure robustness, accuracy, and physical feasibility while providing uncertainty quantification for safe decision-making.

Physics-constrained Motion Prediction (PCMP)  method introduced by~\cite{tumu2023PhysicsConstrainedMotion} 
 combines learning-based intent prediction with physics-constrained trajectory generation and conformal prediction for uncertainty quantification.

Given a sequence of past observations ${O}_i = [{x}_{i,t-l}, \ldots, {x}_{i,t}]$ and environmental context ${C}_i$, the goal is to predict future states ${F}_i = [{x}_{i,t+1}, \ldots, {x}_{i,t+n}]$.
PCMP decomposes this into predicting control inputs ${U}_i$, like steering and acceleration, using a learned intent prediction function 
${\hat{U}}_i = g({O}_i, {C}_i)$
The predicted control inputs are then used to generate the future trajectory using a physics-based model
\begin{equation}
    {\hat{x}}_{i,j+1} = {\hat{x}}_{i,j} + \int_0^{t_s} f({\hat{x}}_{i,j}, {\hat{u}}_{i,j}) \,dt,
\end{equation}
where $t_s$ is the sampling interval.
This guarantees that all predicted trajectories adhere to the robot's physical limits (e.g., steering and acceleration constraints).
The training objective is to minimize the L1 loss between predicted and true trajectories, using an increasing horizon curriculum $h$
\begin{equation}
    \mathcal{L}_h({F}_i, {\hat{F}}_i) = \frac{1}{h} \sum_{j=t+1}^{t+h} \|\lambda \cdot ({x}_{i,j} - \hat{{x}}_{i,j})\|_1,
\end{equation}
where the weight vector $\lambda$ emphasizes key state dimensions.


\subsection{Uncertainty Quantification}
Uncertainty quantification (UQ) is crucial for physics-informed learning because it helps assess the reliability and precision of models that integrate physical laws with data~\cite{yang2019adversarial,zhang2019quantifying}. Taking into account uncertainties in measurements, model parameters, and approximations, the UQ provides information on the confidence of predictions. UQ not only enhances the interpretability of the model but can also be used for reliable decision-making~\cite{abdar2021review} and the design of robust model-based controllers~\cite{hewing2020learning}, ensuring that the learning process remains grounded in both empirical evidence and physical realism. Although there exist multiple approaches for UQ in data-driven models, especially Gaussian processes (GPs) became popular for UQ in physics-informed methods~\cite{long2022autoip,karniadakis2021physics}.

A GP is a nonparametric model widely used in machine learning to perform regression and prediction tasks~\cite{williams1995gaussian}. It can be expressed as a stochastic process $\mathcal{GP}(m_{\mathrm{GP}}(x), k(x,x^\prime))$ on some set $\mathcal{X}$, where any finite collection of points $x^1,\ldots,x^L\in\mathcal{X}$ follows a multivariate Gaussian distribution, fully defined by a mean function $m_{\mathrm{GP}}$ and covariance (kernel) function $k$. Despite many beneficial properties, GPs are, in particular, interesting for physics-informed learning, as they are closed under linear operators~\cite{williams2006gaussian}. Thus, if $\mathcal{L}$ is a linear operator on samples of a GP $f\sim \mathcal{GP}(m_{\mathrm{GP}}(x), k(x,x^\prime))$, then $\mathcal{L}f$ is still a GP\footnote{Assuming the existence of $\mathcal{L}f$, e.g., if $\mathcal{L}$ is the derivative operator, the samples have to be differentiable.}. This property has been used to encode physical prior knowledge to adhere to the Gauss principle of least constraint~\cite{geist2020learning}, to make predictions physically consistent to Euler-Lagrange systems~\cite{evangelisti2022physically}, and port-Hamiltonian systems (PHS)~\cite{9992733}. In the following, we show in detail how GPs can be combined with PHS to achieve physical-consistent predictions with uncertainty quantification.

\paragraph{Gaussian Process port-Hamiltonian Systems (GP-PHS)}
 are probabilistic models for learning partially unknown PHS based on state measurements.  Recall the dynamics of a PHS defined by~\eqref{pHS}.

\begin{figure}
    \centering
\centering\includegraphics[width=0.8\columnwidth]{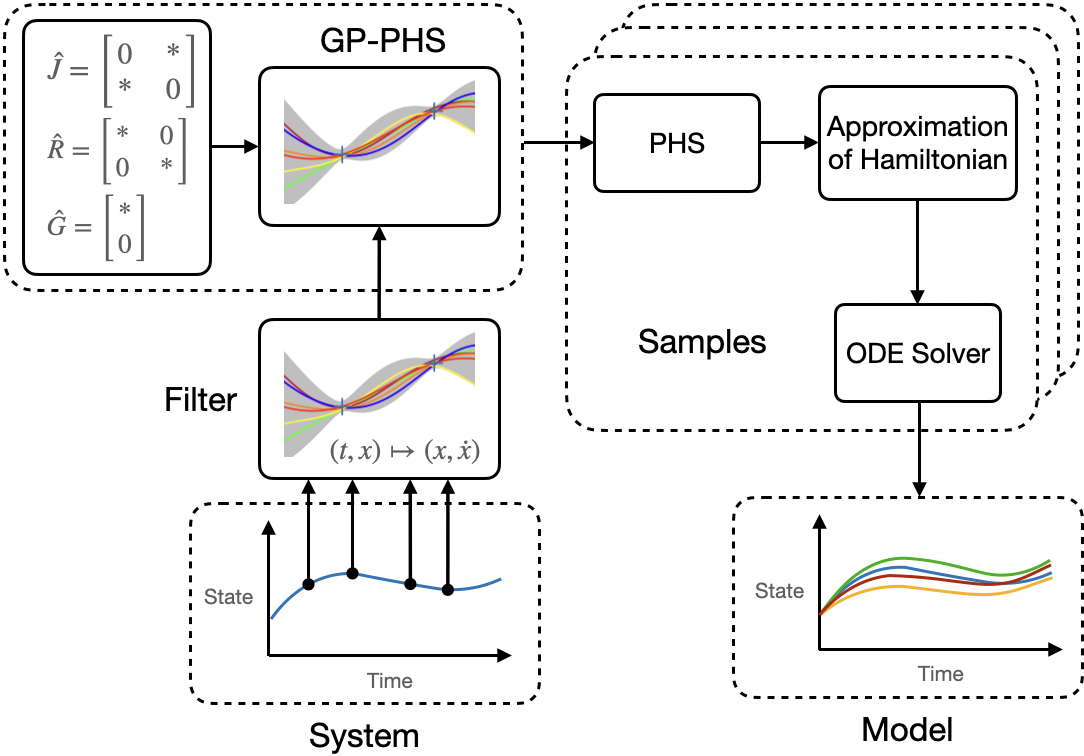}
\caption{\footnotesize Gaussian Process port-Hamiltonian systems enable physically consistent predictions with uncertainty quantification.\label{fig:gpphs}}
\end{figure}

Now, the main idea is to model the unknown Hamiltonian with a GP while treating the parametric uncertainties in $J,R$, and $G$ as hyperparameters, see Fig.~\ref{fig:gpphs}.
By leveraging that GPs are closed under affine operations, the dynamics of a PHS is integrated into the GP by
\begin{align}
    \dot{x}&\sim \mathcal{GP}({\hat G}(x\mid{\varphi}_G)u,k_{phs}(x,x^\prime)),\label{for:gpphs}
\end{align}
where the new kernel function $k_{phs}$ is given by
\begin{align*}
    k_{phs}(x,x^\prime)&=\sigma_f^2\hat{J}_R(x\mid {\varphi}_J,{\varphi}_R)\Pi(x,x^\prime)\hat{J}_R^\top(x^\prime\mid {\varphi}_J,{\varphi}_R)\notag\\
    \Pi_{i,j}(x,x^\prime) &= \frac{\partial }{\partial z_i \partial z_j}\exp(-\|z- z^\prime\|_{\Lambda}^2)\Big\vert_{z=x,z^\prime=x^\prime}
\end{align*}
with the Hessian $\Pi$ of the squared exponential kernel, see~\cite{williams2006gaussian}, and estimates $\hat G,\hat{J}_R=\hat J-\hat R$ . Thus, the dynamics~\eqref{for:gpphs} describes a prior distribution over PHS. We start the training of the GP-PHS by using the collected dataset of timestamps $\{t_i\}_{i=1}^N$ and noisy state observations with inputs $\{\tilde x(t_i),{u}(t_i)\}_{i=1}^N$ of~\eqref{pHS} in a filter to create a dataset consisting of pairs of states $ X=[x(t_1),\ldots,x(t_N)]$ and state derivatives $\dot{X}=[\dot{x}(t_1),\ldots,\dot{x}(t_N)]$. 
Then, the unknown (hyper)parameters $\varphi=[\sigma_f,\Lambda,\varphi_G,\varphi_J,\varphi_R]$ can be optimized by minimizing the negative log marginal likelihood $-\log \mathrm{p}(\dot{X}\vert \varphi,X)\sim\dot{X}_0^\top K_{phs}^{-1} \dot{X}_0+\log\vert K_{phs} \vert$, with the mean-adjusted output data  $\dot{X}_0=[[\dot{x}(t_1)-\hat{G}{u}(t_1)]^\top,\ldots,[\dot{x}(t_{N})-\hat{G}{u}(t_{N})]^\top]^\top$. Once the GP model is trained, we can compute the posterior distribution using the joint distribution with mean-adjusted output data $\dot{X}_0$ at a test states $x^*$
\begin{align}
    \begin{bmatrix}\dot{X}_0\\ \dot{x} \end{bmatrix}\!=\!\mathcal{N}\left({0},\begin{bmatrix}K_{phs} & k_{phs}(X,x^*)\\k_{phs}(X,x^*)^\top & k_{phs}(x^*,x^*)\end{bmatrix}\right)\notag.
\end{align}
Analogously to the vanilla GP regression, the posterior distribution is then fully defined by the mean $\mu\left(\dot{x}\!\mid\!x^{*}, D\right)$ and the variance $\Sigma\left(\dot{x}\!\mid\!x^{*}, D\right)$. Finally, we draw samples from the posterior distribution, where each sample represents a possible PHS given the data set $D$, and compute the solution of each sample using a standard ODE solver. The approach has been extended to enable physically consistent learning of switching systems~\cite{beckers2023learning} and PDE systems~\cite{tan2024physics} with unknown dynamics, and has been applied to design robust, passivity-based controllers~\cite{beckers2023data}.

\paragraph{Uncertainty Quantification with Conformal Prediction in Physics-constrained Motion Prediction (PCMP)}
To quantify prediction uncertainty in PCMP, we can employ conformalized quantile regression (CQR) \cite{angelopoulos_gentle_2022,romano_conformalized_2019}, which constructs statistically valid prediction regions.
A scoring function $s$ measures the deviation between predicted and observed states.
Two novel tailored scoring functions are proposed in PCMP.
In the {Rotated Rectangle Region} method, each predicted state is transformed into the robot's local coordinate frame, aligned with its heading.
The scoring function computes the deviation along the local $x$ and $y$ axes, preserving directional uncertainty. 
In the {Frenet Region} method, where map information is available, errors are defined in the Frenet frame, following the curve of the driving track, where progress along the centerline 
and lateral deviation 
are used. 
These robot-aware prediction regions improve interpretability and ensure that the uncertainty bounds respect road geometry.

Using the above scoring functions, the prediction region is constructed by the CQR method.
The non-conformity score at each timestep is
$R_i = \max \{ q_\text{low} - s_i, s_i - q_\text{high} \}$
where $q_\text{low}$ and $q_\text{high}$ are empirical quantiles from a calibration set.
The final prediction region at confidence level $1 - \delta$ is $
P_{t} = [q_\text{low} - E_{1-\delta}, \; q_\text{high} + E_{1-\delta}]$
where $E_{1-\delta}$ is a quantile of the non-conformity scores.
The true future state lies in $P_t$ with at least $1 - \delta$ probability, even under multi-step predictions.
As the result of uncertainty quantification, these prediction regions are used to ensure the (probabilistic) safety of risk-aware decision-making, such as autonomous robot control or driving algorithms.

\section{Safe Learning for Control}

Ensuring the safety of learned components in control is critical in applications such as autonomous driving and human-robot interaction. In such domains, even small control errors can have severe consequences, necessitating integrated safety mechanisms. As machine learning becomes more prevalent in control systems, maintaining safety and stability remains a major challenge. Unlike traditional controllers, learned policies (e.g., neural networks or adaptive policies) can exhibit unpredictable behavior if unconstrained, underscoring the need for methods that enforce safe behavior at all times.

This section examines how learning can be integrated into model predictive control (MPC) while preserving physical consistency and safety. We cover recent advances in predictive safety filters, which ensure constraint satisfaction under model uncertainty, and explore control barrier and Lyapunov-based methods offering formal safety and stability guarantees. Finally, we discuss Neural Lyapunov Control, which co-designs a policy and a Lyapunov certificate to ensure stability.

\subsection{Learning-based Model Predictive Control and Predictive Safety Filters}
MPC is based on the repeated solution of an optimal control problem (in continuous or discrete time), and the application of the first part of the optimal input before new state information becomes available \cite{rawlings2017model,lucia2016predictive,Diehl2009,MAEDER20092214,Bemporad2002}. For simplicity of presentation, consider  discrete-time dynamical systems given by
\begin{equation} 
\label{eqn:discrete_system_general}
    x_{k+1} = f(x_k, u_k).
\end{equation}
Here $x_k \in \mathbb{R}^{n_\text{x}}$ denote the system states, $u_k \in \mathbb{R}^{n_\text{u}}$ are the system inputs, $f: \mathbb{R}^{n_\text{x}} \times \mathbb{R}^{n_\text{u}} \to \mathbb{R}^{n_\text{x}}$ is the dynamics function, and $k \in \mathbb{N}_0$ is the discrete time index. 

At every discrete time index $k$ and the measurement of the current state $x_k$, one solves the optimal control problem 
\begin{mini!}
    {\mathbf{\hat{u}}_k }  { J = \left\{ \sum_{i=0}^{N-1} \ell({\hat x}_{i \mid k}, {\hat u}_{i \mid k}) + E({\hat x}_{N \mid k}) \! \right\}\label{eqn:mpc_ocp_cost}}{\label{eqn:mpc_ocp}}{}
    \addConstraint{\forall i}{\in \{0, 1, \dots, N-1\}: \notag}{}
    \addConstraint{}{\hat x_{i+1\mid k} = \hat f(\hat x_{i\mid k}, \hat u_{i\mid k}), \ \hat x_{0 \mid k} = x_k,}{\label{eqn:mpc_ocp_model}}
    \addConstraint{}{\hat x_{i \mid k} \in \mathcal{X}, \ {\hat u}_{i \mid k} \in \mathcal{U}, \ \hat x_{N \mid k} \in \mathcal{E}.}{\label{eqn:mpc_ocp_constraints}}
\end{mini!}
Here, $\hat{\cdot}_{i\mid k}$ denotes the model-based $i$-step ahead prediction at time index $k$.
$\hat{f}_ : \mathbb{R}^{n_\text{x}} \times \mathbb{R}^{n_\text{u}} \to \mathbb{R}^{n_\text{x}}, \quad (x, u) \mapsto \hat{f}_\theta(x, u)$ is the prediction model.
The length of the prediction horizon is $N \in \mathbb{N}$, $N < \infty$ and $\ell: \mathbb{R}^{n_\text{x}} \times \mathbb{R}^{n_\text{u}} \to \mathbb{R},$ and $E : \mathbb{R}^{n_\text{x}} \to \mathbb{R},$ are the  stage and terminal cost functions, respectively.
The constraints \eqref{eqn:mpc_ocp_constraints} are comprised of the state, input, and terminal sets $\mathcal{X}_\theta \subset \mathbb{R}^{n_\text{x}}$, $\mathcal{U} \subset \mathbb{R}^{n_\text{u}}$, and $\mathcal{E} \subset \mathbb{R}^{n_\text{x}}$, respectively.
Minimizing the cost over the control input sequence $\mathbf{\hat{u}}_k(x_k)=[\hat u_{0 \mid k}(x_k),\dots,\hat u_{N-1 \mid k}(x_k; \theta)]$ results in the optimal input sequence $\mathbf{\hat{u}}_k^*(x_k)$, of which only the first element is applied to system \eqref{eqn:discrete_system_general}. 
Subsequently, the optimal control problem \eqref{eqn:mpc_ocp} is solved again at all following time indices $k$.
Consequently, the  control policy is given by $u_k = \hat u_{0 \mid k}^*(x_k)$.

If the model perfectly captures the real system dynamics, stability and constraint satisfaction—ensuring safety—can be guaranteed, provided the problem is initially feasible and the cost function, along with the terminal cost, is properly designed \cite{rawlings2017model}. However, in the presence of model uncertainty or external disturbances, robust, set-based, and stochastic MPC approaches have been developed to maintain stability and enforce constraints. While these methods enhance reliability under uncertainty, they often introduce conservatism, potentially limiting performance \cite{bemporad1999robust,mesbah2016stochastic,mayne2016robust,kohler2020computationally,mayne2009robust}.

While robust and stochastic MPC approaches provide safety guarantees under uncertainty, they often lead to conservative behavior. This conservatism arises because these methods typically assume worst-case disturbances or use probabilistic bounds, resulting in overly cautious control actions that limit performance. Set-based approaches, such as tube-based MPC \cite{mayne2009robust}, ensure constraint satisfaction for all possible disturbances but can unnecessarily restrict the system’s capability. Similarly, scenario-based and stochastic MPC \cite{mesbah2016stochastic} formulations must balance constraint satisfaction probabilities with tractability, leading to suboptimal performance in many practical applications.

Learning-based techniques offer a promising path to reduce this conservatism \cite{mesbah2022fusion,hewing2020learning} by adjusting the model, cost function, and constraints based on measured data. However, integrating learning-based components introduces challenges in maintaining feasibility, stability, and robustness. Addressing these challenges has led to a growing focus on ensuring that the learned models and controllers respect physical constraints and stability criteria \cite{brunke2022safe,hewing2020learning,hirt2024safe,Koller2022,Thananjeyan2020}. 

Basically, any component of the MPC formulation \eqref{eqn:mpc_ocp} can be learned, including the cost functions $\ell$, $E$ \cite{jain2021optimal,hirt2024stability,Amos2018}, references and disturbances\cite{rosolia2017learning,matschek2020learninga} the constraints $\mathcal{X}$, $\mathcal{U}$ \footnote{We considered for simplicity set based, deterministic constraints. Often the constraints are given in form of inequalities or in form of probabilistic conditions.} \cite{wabersich2021soft,pfefferkorn2022safe}, or the system model $\hat f$, see \cite{hewing2020learning} for a detailed discussion and classification. Furthermore, one can learn the MPC feedback law itself \cite{karg2020efficient,gonzalez2023neural,rose2023learning,alsmeier2024imitation,DRGONA202280,Drgona2024_DPC,hose2024parameter,hertneck2018learning} or utilize MPC as a safety layer/filter for learning-based controllers \cite{brunke2022safe,mesbah2022fusion}.
Most of these approaches relay on suitably tailored robust predictive control formulations, to guarantee safety and stability, or an sampling based validation of the closed-loop properties.

\paragraph{Learning the Cost Function or Constraints via Differentiable MPC}
A key challenge in safe learning for control is adapting the cost function or constraints in MPC to unknown or evolving system dynamics while maintaining stability and safety guarantees. The work on Differentiable MPC~\cite{Amos2018} provides a framework where both the cost function and constraints can be learned directly from data, enabling end-to-end training of control policies. 
This requires a slight modification of the MPC problem~\eqref{eqn:mpc_ocp} by parametrizing the
 objective $\ell({\hat x}_{i \mid k}, {\hat u}_{i \mid k}, \theta)$ and dynamics constraints $\hat x_{i+1\mid k} = \hat f(\hat x_{i\mid k}, \hat u_{i\mid k}, \theta)$ with trainable parameters $\theta$.

In this approach, an MPC problem is formulated as a differentiable convex optimization layer~\cite{pmlr-v70-amos17a,NEURIPS2019_9ce3c52f}, where the cost function  and system dynamics model parameters are learned jointly through backpropagation.  
 Specifically, differentiable MPC incorporates a differentiable quadratic optimization solver that enables gradient-based updates of $\theta$. By leveraging automatic differentiation, the gradient of the MPC objective with respect to the learnable parameters can be computed as  
\begin{equation}
    \frac{d J}{d \theta} = \sum_{i=0}^{N-1} \frac{\partial \ell}{\partial \theta} + \sum_{i=0}^{N-1} \lambda_t \frac{\partial \hat f}{\partial \theta},
\end{equation}
where \( \lambda_t \) are the Lagrange multipliers associated with the dynamic system constraints. 

An extension of this framework is learning components of a broad family of differentiable control policies~\cite{pmlr-v120-agrawal20a}. 
Other notable extensions to this work, include differentiable robust MPC~\cite{oshin2024} to handle system uncertainties, or using reinforcement learning to tune differentiable MPC policies~\cite{Zanon2021,Gros2020,romero2025} for provably safe control.
 By integrating differentiable optimization with machine learning, this approach bridges the gap between traditional control theory and modern deep learning techniques, enabling real-time adaptive control in dynamic, uncertain environments.

\paragraph{Predictive Safety Filters (PSF)}
PSF ensure safe operation in control systems by modifying control inputs in real time to prevent constraint violations. These filters use predictive models of system dynamics~\eqref{eqn:discrete_system_general} that can be either physics- or learning-based to anticipate unsafe situations and minimally adjust control actions only when necessary, allowing the nominal controller to operate freely when safety is not at risk. They are particularly useful in learning-based control, where data-driven policies may lack formal guarantees of safety. 

A predictive safety filter solves a constrained optimization problem at each time step
\begin{subequations}
\begin{align}
    \min_{u_t} & \quad \| u_t - u_t^\text{nom} \|^2 \\
    \text{s.t.} & \quad x_{t+1} = f(x_t, u_t), \\
    & \quad h(x_t, u_t) \geq 0, \quad \forall t.
\end{align}
\end{subequations}
where \( u_t^\text{nom} \) is the nominal control input generated by an external policy (e.g., a reinforcement learning controller), and \( h(x_t, u_t) \geq 0 \) defines a safety constraint that must be satisfied at all times. This formulation ensures minimal deviation from the nominal control while enforcing safety.

An effective implementation of predictive safety filters is based on MPC~\cite{WABERSICH2021109597}, which predicts future states over a finite horizon and optimizes control adjustments to ensure safety. Unlike traditional MPC, which is designed for optimal performance, predictive safety filters focus purely on safety enforcement, allowing the primary controller to retain its autonomy whenever possible. 
This approach formulates the MPC safety filtering problem as a quadratic program (QP) that needs to be solved online to guarantee the stability of the closed-loop system~\cite{milios2024}. 
Due to their stability guarantees based on MPC  theory~\cite{pmlr-v211-leeman23a}, predictive safety filters have been applied in safety-critical applications such as autonomous racing~\cite{Tearle2021}.
Extensions to MPC-based predictive safety filters include conformal prediction to provide uncertainty quantification~\cite{Strawn2023}, relaxed formulations with soft-constrained control barrier functions~\cite{Wabersich2023}, extensions to nonlinear systems dynamics~\cite{DIDIER2024200}, or event-triggered mechanisms for reducing the computational requirements~\cite{cortez2024}.

Despite their advantages, MPC-based safety filters require QP solvers capable of handling the online computation of constrained control adjustments. Recent advancements in  fast quadratic programming solvers~\cite{osqp} have improved the scalability of these methods, enabling their application in complex, high-dimensional systems. By integrating predictive safety filters with learning-based controllers, it is possible to achieve both adaptability and formal safety guarantees, enhancing the robustness of real-world control systems.

\subsection{Control Barrier and Control Lyapunov Functions}





Lyapunov Functions (LF), Control Lyapunov Functions (CLFs), and Control Barrier Functions (CBFs) are fundamental tools in ensuring stability and safety in control systems, particularly when integrating learning-based or data-driven models with formal guarantees~\cite{Ames2017,Ames2019,Taylor2020,Dawson2023,Brunke2022,Wang2025,Wabersich2023,NIPS2017_766ebcd5,Prajna2004,Prajna2007,Barry2012}.
Let's assume a dynamical system in control-affine form
\begin{equation}
\label{eq:ctrl_affine_dynamics}
\dot{x} = f(x) + g(x) u.
\end{equation}
A CLF is used to ensure stability of~\eqref{eq:ctrl_affine_dynamics} by providing a scalar function that decreases along system trajectories under an appropriate control input. Specifically, a continuously differentiable function \( V(x) \) is a CLF if there exist positive constants \( c_1, c_2 > 0 \) such that  
$    c_1 \| x \|^2 \leq V(x) \leq c_2 \| x \|^2, \quad \forall x \in \mathbb{R}^n$,
and for all nonzero \( x \), there exists a control input \( u \) such that  
\begin{equation}
    \inf_{u} \left[ \frac{\partial V}{\partial x} (f(x) + g(x) u) \right] \leq -\alpha(V(x)),
\end{equation}
for some positive class-\(\mathcal{K}\) function \( \alpha(\cdot) \). This condition ensures that an appropriate control input can always be found to drive the system towards equilibrium, guaranteeing asymptotic stability.

On the other hand, a CBF is designed to enforce safety constraints by ensuring that the system remains within a specified safe set. If a function \( h(x) \) defines the safe set as  $\{ x \mid h(x) \geq 0 \}$,
then a valid control input must satisfy  
\begin{equation}
    \sup_{u} \left[ \frac{\partial h}{\partial x} (f(x) + g(x) u) \right] \geq -\alpha(h(x)),
\end{equation}
ensuring that trajectories do not exit the safe region. In practical implementations, CBFs and CLFs are often combined in a quadratic program framework to balance stability and safety objectives, particularly in real-time control of robotic systems, autonomous vehicles, and other safety-critical applications. These techniques are increasingly integrated with machine learning models to provide certifiable safety guarantees in data-driven controllers, ensuring that learned policies respect both stability and safety constraints in uncertain environments.

In the following two paragraphs, we provide two related examples for data-driven neural Lyapunov function candidates for stability verification of autonomous and closed-loop controlled dynamical systems with learning-based controllers.

 \paragraph{Learning Neural Lyapunov Functions}
The Lyapunov function (LF) can be interpreted as a scalar-valued function of potential energy stored in the system. LF is a crucial concept in the control theory and stability theory of dynamical systems. For certain classes of ODEs, the existence of LF is a necessary and sufficient condition for stability. However, obtaining LF analytically for a general class of systems is not available, while existing computational methods, such as the Sum of Squares method~\cite{Papachristodoulou2025}, are intractable for larger systems. Approaches for learning neural LF candidates from dynamical system data~\cite{Deka2022,LyapunovNN2018,NEURIPS2019_0a4bbced,pmlr-v87-richards18a,pmlr-v162-rodriguez22a,Chen2021} presented a tractable alternative. 

Let's consider the problem of learning the LF that can be used to certify the stability of a dynamical system based on the observation of its trajectories. 
Specifically, lets assume autonomous system dynamics given by $\dot{x} = f(x); \,\,\, x \in R^n $, with available dataset of state observations $X = [x^i_1, ..., x^i_N]$, $i = 1,...,m$.  
Where $N$ defines the length of the rollout, and $m$ defines the number of sampled trajectories. The objective is to find a LF candidate $V_{\theta}(x): R^n \to R; \,\,\, V_{\theta}(0) = 0; \,\,\,  V_{\theta}(x) >0, \forall x \neq 0 $, parametrized by ${\theta}$, that satisfies the discrere-time Lyapunov stability criterion  $V_{\theta}(x_{k+1}) - V_{\theta}(x_k) < 0$  for the given dataset of trajectories $X$.

We chose the neural LF candidate as presented in~\cite{NEURIPS2019_0a4bbced}.
The architecture is based on the 
 input convex neural network (ICNN)~\cite{ICNN2017} followed by a positive-definite layer.
 The ICNN with $L$ layers is defined as
 \begin{subequations}
\begin{align}
   & z_1 = \sigma_0(W_0 x + b_0) \\
   & z_{i+1} = \sigma_i(U_i z_i + W_i x + b_i), \,\,\, i  \in \mathbb{N}_1^{L-1} \\
  & g_{\theta}(x) =  z_k  
\end{align}
 \end{subequations}
Where $W_i, b_i $  are real-valued weights and biases, respectively, while $U_i$ are positive weights, 
and $\sigma_i$ are convex, monotonically non-decreasing non-linear activations of the $i$-th layer.
The trainable parameters of the ICNN architecture are given by $\theta = \{ W_0, b_0, W_i, b_i, U_i \}, i \in \mathbb{N}_1^{L-1} $.
This ICNN layer guarantees that there are no local minima in the LF candidate $V_{\theta}(x)$.
The ICNN is followed by a positive-definite layer 
 \begin{equation}
    V_{\theta}(x) = \sigma_{k+1}(g_{\theta}(x) - g_{\theta}(0)) + \epsilon ||x||^2_2,
 \end{equation}
that acts as a coordinate transformation enforcing the  condition $V_{\theta}(0) = 0$.

The general data-driven approach is to
use stochastic gradient descent to optimize the parameters $\theta$ of the neural LF candidate $V_{\theta}(x)$ over the dataset of the state trajectories $X$ using the physics-informed neural network (PINN)~\cite{raissi2019physics} approach of penalizing the expected risk of constraints violations given by the loss function
 \begin{equation}
 \label{eq:Lyapunov_loss}
  \mathcal{L}(X)  = \sum_{i=1}^m \sum_{k=1}^{N-1} || \max \{0, V_{\theta}(x^i_{k+1}) - V_{\theta}(x^i_k)\} ||_2^2.
 \end{equation}

 The disadvantage of the PINN approach with a soft-constrained loss function~\eqref{eq:Lyapunov_loss} is that it does not satisfy the Lyapunov stability criterion. To alleviate this issue in the context of system identification and learning-based controls, authors have proposed additional projection layers to correct the dynamics of the modeled system~\cite{NEURIPS2019_0a4bbced}, to guarantee Lyapunov stability~\cite{donti2021enforcing}, or enforce convex constraints~\cite{Chen2021}.

The open-source code demonstrating this tutorial example on a specific dynamical system can be found in the NeuroMANCER library~\cite{Neuromancer2023}.

\paragraph{Physics-informed Neural Lyapunov Control}
Beyond learning a Lyapunov certificate for a given autonomous system, one can jointly design a control policy and its corresponding Lyapunov stability certificate. 
Some of the existing works on this topic include~\cite{mittal2020neural,chang2019neural,dawson22a,LyapunovDPC22,Wei2023,Wang2024,Liu2024,liu2024formal,pmlr-v235-meng24b}.

\begin{figure}
\centering
\includegraphics[width=0.3\textwidth]{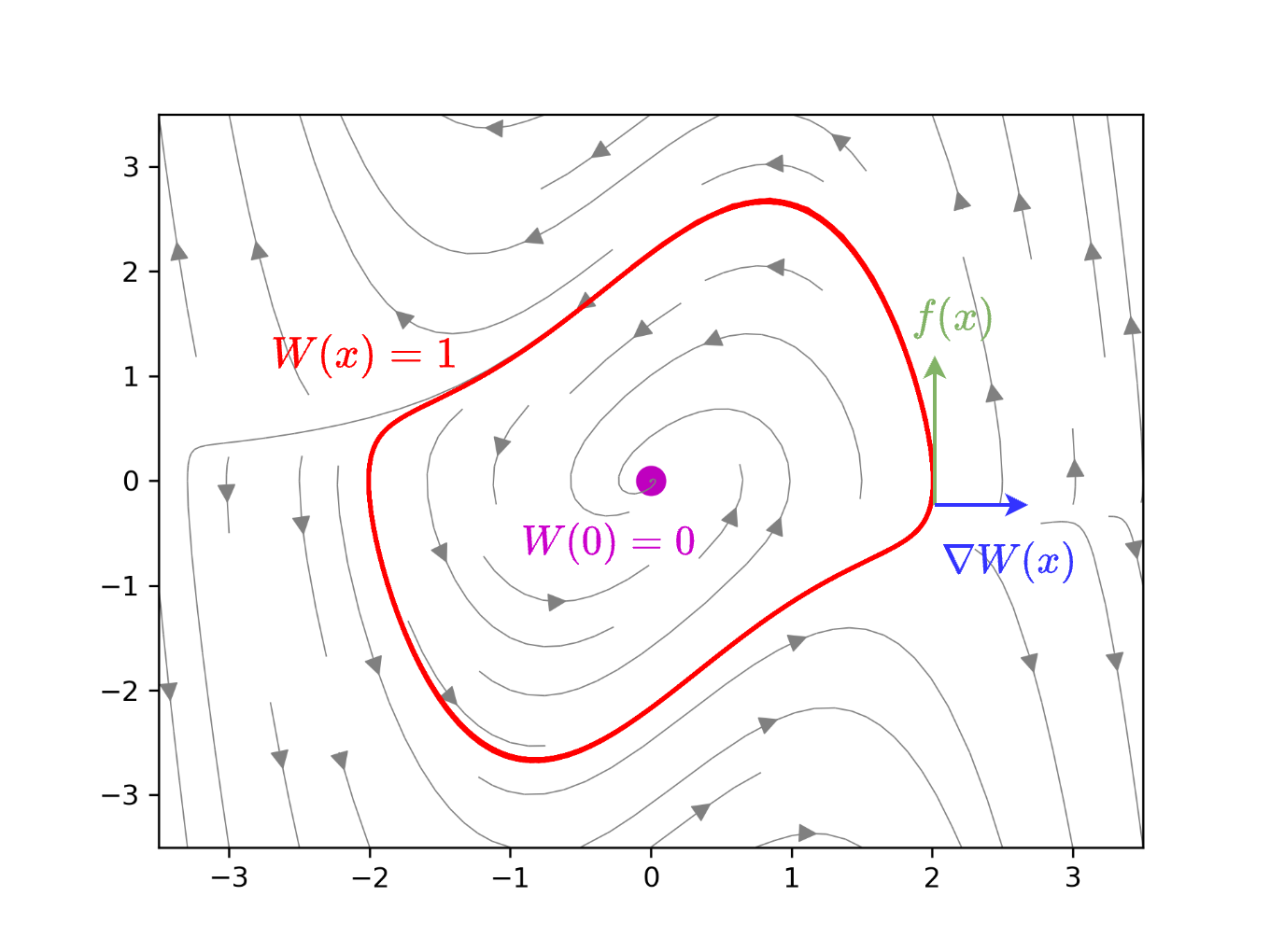}
\caption{The level set $\{x : W(x)=1\}$ obtained by the Zubov PDE characterizes the boundary of the true DoA.}
\label{fig:zubov}
\end{figure}

\begin{figure}
     \centering
     \begin{subfigure}[b]{0.23\textwidth}
         \includegraphics[width=\textwidth]{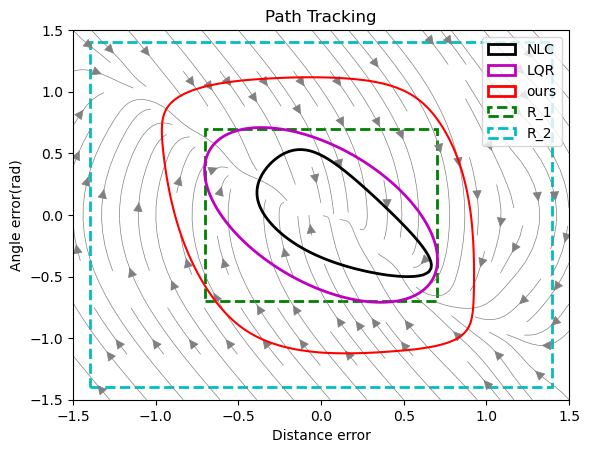}
         \caption{$W_{\theta}(x)$ learned for bicycle tracking system}
         \label{fig:tracking_3d}
     \end{subfigure}
     \begin{subfigure}[b]{0.23\textwidth}
         \includegraphics[width=\textwidth]{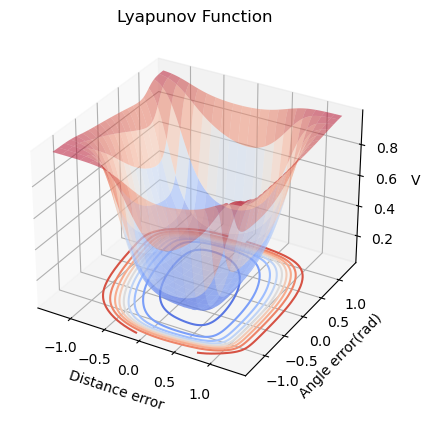}
         \caption{Verified DoA and vector field $\dot{x} = f(x, \pi_\gamma(x))$ }
         \label{fig:tracking_2d}
     \end{subfigure}
        \caption{Bicycle tracking system.}
        \label{fig:tracking}
\end{figure}

Consider the nonlinear system  $\dot{x} = f(x, u)$,  where \( (x,u) \in \mathcal{D} \times \mathcal{U} \) is the state-control pair. The goal is to design or learn a control policy \( \pi_{\gamma}(\cdot) \) that stabilizes an equilibrium while optimizing a performance metric, such as maximizing the region of attraction (RoA), which serves as a robustness margin against uncertainties.
A common learning-based approach parameterizes both the Lyapunov function \( V_\theta \) and controller \( \pi_{\gamma} \) using neural networks. Stability is enforced by minimizing the expected violation of Lyapunov conditions
\[
\min_{\theta, \gamma} E_{x \sim \rho} \bigg[ 
    V_\theta(0)^2 + (-V_\theta(x))_{+} + \big(\nabla_x V_\theta(x)^\top f(x, \pi_\gamma(x))\big)_{+}
\bigg].
\]
However, this optimization can admit trivial solutions, such as \( V_{\theta}(x) = 0 \) for all \( x \in \mathcal{D} \). Regularizations such as  $( \|x\|_2 - \alpha V_\theta(x) )^2$ can be used to encourage large RoAs \cite{chang2019neural}, though poor choices may misalign the learned Lyapunov level sets with the true RoA. To address this, \cite{LIU2025112193,Wang2024} propose a physics-informed loss based on Zubov’s PDE, which directly captures the RoA. The PDE,
\[
\nabla W(x)^\top f(x,\pi_{\gamma}(x)) + \Psi(x)(1 - W(x)) = 0,
\]
defines the Zubov function \( W(x) \in (0,1) \), with \( \Psi(x) \) as a positive definite function. Intuitively, this ensures that the Lie derivative of \( W \) is zero on the RoA boundary, precisely identifying the domain of attraction (see Figure \ref{fig:zubov} for an illustration).
Guided by Zubov's PDE, the learned Lyapunov function is depicted in Figure \ref{fig:tracking_3d}. The verified region of attraction and the vector field of the learned controller are shown in Figure \ref{fig:tracking_2d}.

\section{Safety Verification Methods}
 Safety verification methods aim to formally or empirically certify that a system operates within predefined safety constraints under all possible disturbances and uncertainties. These methods broadly fall into three categories: reachability analysis methods~\cite{GARG2024100948,Zhang2020}, constrained optimization-based methods~\cite{Fazlyab2022,Zhang2022}, and simulation and sampling-based approaches~\cite{liebenwein2018sampling,chakrabarty2018data,shen2022model}. Reachability analysis constructs forward or backward reachable sets to determine whether unsafe states can be reached from an initial condition, providing rigorous guarantees of safety but often facing scalability challenges in high-dimensional systems. Constrained optimization-based approaches formulate safety verification as an optimization problem where system constraints, safety conditions, and physical limitations are explicitly incorporated. The goal is to determine whether a system can violate predefined safety conditions by solving an optimization problem that seeks counterexamples or guarantees safe operation.
Simulation and sampling-based approaches, including Monte Carlo simulations and probabilistic verification, offer a scalable alternative by empirically evaluating system behavior over a large number of sampled scenarios, often at the cost of completeness. Together, these techniques form a diverse toolkit for verifying and enforcing safety in complex, uncertain environments, striking a balance between computational feasibility and formal guarantees.

\subsection{Reachability Analysis Methods} 
In safety-critical systems, reachable set analysis determines all possible states a dynamical system can reach from a given set of initial conditions and inputs. Safety verification often requires ensuring that the reachable set never intersects unsafe states. However, computing exact reachable sets is generally intractable, except in special cases. Existing approaches for reachability analysis fall into two broad categories: set propagation techniques and Hamilton-Jacobi reachability.

\paragraph{Set Propagation Techniques} Given the system  
$x^{k+1} = f(x^k), \quad x^0 \in \mathcal{X}^0$,
where \(\mathcal{X}^0\) is a bounded set of initial conditions, the reachable set at time \( k+1 \) is  
\[
\mathcal{X}^{k+1} = f(\mathcal{X}^k) = \{f(x) \mid x \in \mathcal{X}^k\}.
\]
Since exact computation is generally intractable, reachable sets are typically over-approximated using {template sets} \((\bar{\mathcal{X}}_k)_{k\geq 0}\), where \(\bar{\mathcal{X}}^0 = \mathcal{X}^0\) and each step ensures \( f(\bar{\mathcal{X}}^k) \subseteq \bar{\mathcal{X}}^{k+1} \) via a {set propagation algorithm} often relying on convex relaxations. If these sets avoid unsafe regions, safety is guaranteed. 

Convex relaxations improve computational tractability but introduce conservatism, particularly when the actual reachable set is nonconvex or irregularly shaped. The resulting over-approximation errors can accumulate over time—a phenomenon known as the wrapping effect \cite{neumaier1993wrapping}—leading to increasingly conservative bounds. 

Conservatism due to nonconvexity can be reduced by employing tighter convex relaxations that aim to approximate the convex hull of the reachable set, partitioning input sets to refine the resolution or selecting templates that better match the geometry of the reachable set. The choice of template sets, such as ellipsoids or polyhedra, significantly impacts accuracy. A poor match between the template shape and the true reachable set leads to shape mismatch error, even with tight convex relaxations.

Dynamic templates can help mitigate these errors by adapting to the geometry of $f$. Oriented hyperrectangles, obtained via principal component analysis (PCA) on sampled trajectories \cite{stursberg2003efficient}, are computationally efficient but can still be overly conservative. More sophisticated methods exist, such as those in \cite{bogomolov2017counterexample}, which solve convex optimization problems to determine template polytope directions, or in \cite{ben2012reachability}, where a first-order Taylor approximation of $f$(assuming differentiability) is used to guide template adaptation. A more recent approach in \cite{entesari2023automated} leverages a ReLU network to dynamically adjust polytopic templates, offering a data-driven way to better approximate the reachable set.

\paragraph{Hamilton Jacobi Reachability}
Hamilton–Jacobi (HJ) reachability is a powerful method for computing reachable sets in dynamical systems by leveraging optimal control theory and level-set methods. It formulates safety verification as a Hamilton–Jacobi–Isaacs partial differential equation (HJI-PDE)~\cite{mitchell2005time}, where the value function 
$V(x,t)$ defines the boundary of the reachable set as its zero level-set. By solving this PDE forward or backward in time, HJ reachability provides precise characterizations of system behaviors, accounting for nonlinear dynamics and adversarial disturbances. This rigorous approach makes HJ reachability a gold standard in safety verification, particularly for applications requiring guaranteed safety under worst-case uncertainties. However, despite its accuracy, the method suffers from the curse of dimensionality, as solving the HJI-PDE becomes computationally intractable for high-dimensional systems, limiting its scalability to complex real-world applications. Recent advancements, including neural approximations~\cite{Hofgard2024ConvergenceGF,DARBON2021109907,darbon2020} and decomposition techniques~\cite{CHOW2019376,Chen2017}, aim to mitigate these limitations and extend the applicability of HJ reachability to large-scale dynamical systems.


\subsection{Constrained Optimization-based Methods} 


Many verification problems can be posed as proving the positivity of a function over a specified domain $\min_{x\in X} f(x) \ge 0$. Common examples include demonstrating the existence of a Lyapunov function, computing an invariant region in which constraints are satisfied, and certifying a maximum error for an approximate control computation over a region of interest. Such problems have long been studied in the control literature, and the ability to solve these challenging problems depends very heavily on the properties of the system, control law and property being verified (the function $f$), as well as the structure of the domain $X$.

\paragraph{Verificaiton via Mixed-integer Linear Programming}
One approach that can be used for a number of important verification problems is to map the positivity verification problem onto a mixed-integer linear programming structure (MILP), which is then solvable to global optimality using standard tools, thereby providing rigorous guarantees for a variety of physical properties, as well as safety and stability. Problems that have such a mapping are dubbed `MILP-representable verification problems' \cite{schwan2023}.

At the core of these approaches is the concept of MILP-representable functions. Such functions \( \psi: X \rightarrow U \) have graphs that can be exactly characterized by linear inequalities and equalities involving continuous and binary variables. Formally, a function \( \psi : \mathbb{R}^n \mapsto \mathbb{R}^m \) is called MILP-representable if there exists a polyhedral set \( P \subseteq \mathbb{R}^{n+m+c+b} \) such that
\begin{equation}
    (x,u) \in \text{gr}(\psi) \Leftrightarrow \exists p \in \mathbb{R}^c, \beta \in \{0,1\}^b: (x,u,p,\beta) \in P.
\end{equation}
One can show that the set of MILP-representable functions are dense in the family of continuous functions on a compact domain with respect to the supremum norm, and that the composition of MILP-representable functions is also MILP-representable. This formulation allows a diverse set of systems and controllers to be modeled exactly, including piecewise-affine systems, the optimal solution map of a parametric quadratic program (such as MPC), or controllers defined by ReLU NNs.
To illustrate, we take the simple example of a controller that computes its input via a fully-connected deep neural network $\phi : \mathbb{R}^n \rightarrow \mathbb{R}$, which is a composition of ReLU layers $\phi = \phi_l \circ \phi_{l-1} \circ \dots \circ \phi_0$, where $\phi_i : \mathbb{R}^{n_i} \rightarrow \mathbb{R}^{m_i}$. The ReLU function is defined as 
\begin{align*}
    \phi_i(z_{i-1}) = ReLU(z_{i-1}) = \max(0, W_iz_{i-1}+b_i),
\end{align*}
for weight matrix $W_i$ and bias vector $b_i$. We define the polyhedron $P_i$ as 
\begin{align*}
    P_i = \left\{
        (z_{i-1},z_i,p,\beta_i)\ \left|\ 
        \begin{aligned}
        |z_i - (W_iz_{i-1}+b_i)| &\le M(1-\beta_i)\\
        0 \le z_i &\le M\beta_i
        \end{aligned}
        \right.\right\}
\end{align*}
where $M$ is a sufficiently large positive constant. We can see that $z_i = \phi_i(z_{i-1})$ if and only if there exists a $\beta_i \in \{0,1\}^{m_i}$ such that $(z_{i-1}, z_i, p, \beta_i) \in P_i$.

Suppose now that we have the simple verification goal of computing the maximum value that the controller may take for any state in the polyhedral set $X$. i.e., we want to solve the verification problem $\max_{x \in X}\phi(x)$. Because the controller is MILP-representable, this problem can be written as a standard MILP~\eqref{eq:MILP}.
\begin{subequations}
\label{eq:MILP}
    \begin{align}
    \max_{x\in X}&\ u\\
    \text{s.t.}&\ 
    \begin{aligned}[t]
        (x, z_1, p_1, \beta_1) &\in P_1\\
        (z_{i-1}, z_i, p_i, \beta_i) &\in P_i\quad \forall i = 2\dots,l-1\\
        (z_{l-1}, u, p_l, \beta_l) &\in P_l\\
        \beta_i &\in {0,1}^{n_i}\quad \forall i = 1\dots,l
    \end{aligned}
\end{align}
\end{subequations}

While we have followed the formalism introduced in \cite{schwan2023} here, there are several variations of this concept in the literature. \cite{sergio2020} captures neural-network controllers in MILP form, before doing a reachability analysis, while \cite{hongkai2020,shaoru2021} represent PWA Lyapunov functions in a similar formalism while executing a learner/verifier pattern. \cite{fabiani2023} combines a MILP representation with a Lipschitz bound to conservatively compute a bound on the approximation error of a neural network. 
Others~\cite{NEURIPS2019_95e1533e,Fazlyab2022,Pauli2022,PauliSDP2022} propose solving a semidefinite program (SDP) to obtain a tight and certifiable upper bound on the global Lipschitz constant of a neural network.

\subsection{Simulation and Sampling-based Methods}

Simulation or sampling-based methods provide a simple and scalable alternative for verification tasks for complex dynamical systems with possibly unknown components. 
Due to their flexibility and practical performance in high-dimensional systems, they have gained popularity for approximating reachable sets~\cite{liebenwein2018sampling,lew2021sampling,lew2022simple,lin2023generating,gruenbacher2022gotube,devonport2020estimating}, invariant sets~\cite{chakrabarty2018data,wang2019data,mulagaleti2021data}, or regions of attraction~\cite{shen2022model,kang2023data}.
In this section, we briefly overview its main components and the types of guarantees that one can achieve.

\paragraph{Sampling-based Reachability Analysis}
In sampling-based reachability analysis, one is given a set $\mathcal{X}$ and map $f:\mathbb{R}^m\rightarrow\mathbb{R}^n$, with the goal of approximating the set
\begin{equation}\label{eq:reach-set}
    \Y=f(\X)=\{f(x):x\in\X\}
\end{equation}
that is reached from $\X$ using samples. Methods proposed in this context often consider the following elements:
\begin{itemize}
    \item A distribution $\mathbb{P}_\X$ to sample points from $\X$.
    \item A number $N$ of i.i.d. samples $\{(x_i, y_i = f(x_i))\}_{i=1}^N$.
    \item A parameter $\epsilon$ quantifying conservativeness.
    \item A hypothesis class $\mathcal{H}\subset 2^{\mathbb{R}^n}$ of candidates set $\hat\Y$.
\end{itemize}
Given these elements, the role of sampling-based algorithms is to construct a set-valued map $\hat\Y_\epsilon^N$ that maps the samples to an estimated set $\hat\Y \in \mathcal{H}$:
\[
\hat\Y_\epsilon^N:\{y_i\}_{i=1}^N \mapsto \hat\Y\in\mathcal{H}.
\]
The choice of these elements determines the outcome of this process, computational complexity, and guarantees.

\textit{Choice of $\Y_\epsilon^N$ and $\mathcal{H}$:} In general, the selection of the hypothesis class $\mathcal{H}$, and as a result, the implementation of $\Y_\epsilon^N$ could be as simple as computing a union balls $\epsilon$-padded points or normed balls~\cite{korostelev2012minimax,rodriguez2016fully,liebenwein2018sampling,gruenbacher2022gotube}, e.g.,
$
\hat\Y_\epsilon^N = \{y_i\}_{i=1}^{N}\oplus B_\epsilon,
$ its convex hull, i.e.,
$\hat\Y_\epsilon^N = \mathrm{conv}\left(\{y_i\}_{i=1}^{N}\right)\oplus B_\epsilon$~\cite{lew2021sampling,lew2022simple}, or given by the $\epsilon$-sublevel set of a neural function approximation $\hat\Y_\epsilon^N = \{y: V_\theta(y)\leq \epsilon\}$~\cite{bansal2021deepreach}, where the function $V_\theta$ is trained using the samples. The complexity of computing the approximation can vary significantly.

\textit{Distribution $\mathbb{P}_\X$:} The sampling distribution $\mathbb{P}_\X$ is often a design parameter. 
Simple yet effective algorithms often choose a fix $\mathbb{P}_\X$, e.g., uniform, whose support covers $\X$ or its boundary $\partial\X$. Alternatively, adaptive sampling approaches, e.g., using active learning~\cite{chakrabarty2018data}, can be further leveraged to improve sample efficiency.

\textit{Formal Guarantees:} Finally, guarantees on the validity and accuracy of the verification process can be given for different choices of $N$, $\epsilon$, and $\mathcal{H}$. These can be succinctly expressed using some notion of set distance $d(P, Q)$. A common example is the Hausdorff distance $$d_H(P, Q)=\max\{\,\adjustlimits\sup_{x\in P} \inf_{y\in Q}d(x,y),\; \adjustlimits\sup_{y\in Q} \inf_{x\in P}d(x,y)\,\}.$$
In some cases, other accuracy metrics are considered, e.g., $d_\mathbb{P_\X}(\hat{\Y}_\epsilon^N, \Y) = \mathbb{P}_\X\left(\hat{\Y}_\epsilon^N\cap \Y\right)$, although they struggle to capture the outer level of mismatch $\hat{\Y}_\epsilon^N$ and $\Y$~\cite{devonport2020estimating}.

\textit{Almost Sure Guarantees:} Taking $N$ to infinity leads to
 \begin{equation}\label{eq:as-guarantee}
     \lim_{N\rightarrow\infty} d_H(\hat \Y_\epsilon^N, \Pi_\mathcal{H}(\Y)) \leq \epsilon \quad \text{a.s.}
 \end{equation}
 where $\Pi_\mathcal{H}(\Y)$ is the best approximation of $\Y$ within $\mathcal{H}$, e.g., $\Pi_{\mathcal{H}}(\Y) = \mathrm{conv}(\Y)$ for convex hull approximations.
When $\epsilon=0$, this leads to consistent estimates~\cite{lew2021sampling}.

\textit{High-probability Guarantees:}
For sufficiently large $N$, e.g., $N \geq \Omega\left(\mathrm{poly}\left(\frac{1}{\epsilon}\right)\mathrm{polylog}\left(\frac{1}{\delta}\right)\right)$, the condition \eqref{eq:as-guarantee} often reduces to guaranteeing
\begin{equation}\label{eq:high-prob-guarantee}
    d_H(\hat \Y_\epsilon^N, \Pi_\mathcal{H}(\Y)) \leq \epsilon
\end{equation}
with probability $1-\delta$~\cite{lew2022simple}, provided some additional assumption is considered on $f$, e.g., Lipschitz. 
Depending $\mathcal{H}$  and the value of $\epsilon$, the corresponding $\hat\Y_\epsilon^N$, can also be guaranteed to either an inner approximation, for $\epsilon=0$, or outer approximation, for $\epsilon>0$ and $N$ large.

\paragraph{Sampling-based Estimation of Invariant Sets} 
Invariant sets, i.e., sets that keep trajectories within a specified region, can be similarly estimated via iterative sampling strategies similar to the ones described above. We briefly discuss here the main differences. In this setting, given a domain $\X$, we are interested in identifying set $\I\subseteq \X$, with the properties that trajectories $x(t)$ satisfying 
\begin{equation}\label{eq:invariance-condition}
x(0)\in \I\implies x(t)\in \I,\quad \forall t\geq0.
\end{equation}
Similarly to the reachability setting, one analogously defines $\mathbb{P}_\X$, number of samples inside and outside $\I$, $N_i$ and $N_o$ with $N=N_i+N_o$, a conservativeness parameter $\epsilon$, and a hypothesis class $\mathcal{H}$, and set estimate $\hat\I_\epsilon^N$. In particular, similar hypothesis classes $\mathcal{H}$ are considered, including unions of balls~\cite{shen2022model}, sub-level sets of neural function approximations~\cite{chakrabarty2018data,kang2023data}, or biased versions of nearest neighbors~\cite{wang2019data}.
However, there are key notable differences in this setting that we highlight next.

\textit{Circular dependence of} $\hat\I_\epsilon^N$:
Given some $x\in\X$, checking condition \eqref{eq:invariance-condition} requires knowledge of $\I$, yet building an estimate $\hat\I_\epsilon^N$  requires the identification of point $x_i\in\I$ and $x_o\not\in\I$. This is usually circumvented by constructing implicit oracles of the condition $\mathcal{O}(x) = \mathbb{1}_{\{x\in\Y\}}$ checking implicit invariance conditions such as convergence to a small ball around a stable equilibrium~\cite{chakrabarty2018data},  finding a time $t^*$ s.t. if $x(t^*)\in\X$, then $x(t)\in\X$ $\forall t\geq t^*$~\cite{wang2019data}, or checking a recurrence condition on $\hat\I_\epsilon^N$,~\cite{shen2022model}, that guarantees the inclusion $\hat\I_\epsilon^N\subseteq \I$.

\textit{Non-uniqueness:} The number of invariant sets within $\X$ may be uncountable, thus introducing a level of indeterminism that makes the problem more challenging and algorithms more difficult to converge. This is often overcome by targeting invariant sets that are maximal in some sense, like domains of attraction of stable equilibrium points~\cite{shen2022model,kang2023data}.

\textit{Formal Guarantees:} Finally, we highlight that the type guarantees focus on quantifying the fraction of the volume of $\hat\I_\epsilon^N$ that is invariant, e.g., stating that with probability $1-\delta$, $d_{\mathbb{P}_\X}(\hat\I_\epsilon^N,\I)=\mathbb{P}_\X(\hat\I_\epsilon^N\cap\I)\leq \epsilon$ as in~\cite{wang2019data}. However, it is guaranteeing $\hat\I_\epsilon^N$ being invariant (with high probability) is generally difficult. At best, guarantees of the form $\hat\I_\epsilon^N\subset \I$ with high probability, or a.s. in the limit $N\rightarrow\infty$ are obtained in practice~\cite{shen2022model}.

\section{Challenges and Opportunities}

The integration of safe physics-informed machine learning (PIML) in dynamical systems and control presents several challenges and opportunities. 

A key challenge in uncertainty quantification (UQ) is reliably estimating uncertainties in learned models and translating them into actionable safety constraints. While methods such as Gaussian processes and Bayesian neural networks provide probabilistic estimates, applying them in safety-critical control remains challenging. Advances in conformal prediction, probabilistic reachability, and risk-aware optimization offer promising solutions. Future work should integrate these UQ methods into real-time safety mechanisms such as safety filters and control barrier functions.

A major challenge in safe learning-based control is balancing conservatism and performance. While robust and stochastic MPC enforce safety, they can be overly conservative and limit performance. Learning-based methods, such as differentiable MPC, offer adaptability by updating constraints and models based on data, but raise open questions regarding stability, robustness, and computational feasibility. Promising directions include adaptive safety filters, robust learning frameworks with real-time uncertainty quantification, and offline pre-training or event-triggered strategies to reduce online computation.

Another major challenge is the safe exploration problem in reinforcement learning (RL), where unrestricted exploration can lead to safety violations. While approaches such as control barrier functions (CBFs), Lyapunov-based stability certificates, and predictive safety filters (PSFs) provide mechanisms for enforcing safety, they can overly constrain learning, limiting policy improvement. A key research opportunity lies in developing efficient RL exploration strategies that balance safety with learning efficiency, potentially leveraging safe Bayesian optimization, risk-sensitive RL, and constrained offline policy optimization methods.

A key challenge in safe learning is scalability. Methods like Hamilton–Jacobi reachability and mixed-integer safety verification often face the curse of dimensionality, limiting their use in high-dimensional systems. While decomposition and neural approximations offer some relief, achieving both accuracy and efficiency remains difficult. Developing scalable verification techniques with strong safety guarantees is a critical direction for future work.

From a broader perspective, there is a need for unified frameworks that integrate multiple safety mechanisms. Current approaches often treat safe learning, safety verification, and robust control as separate problems. A more holistic approach would involve the co-design of learning algorithms, verification techniques, and real-time control strategies, ensuring that learned policies inherently satisfy safety and stability constraints. Such frameworks could leverage advances in differentiable optimization, neural-symbolic reasoning, and hybrid machine learning-control architectures to create adaptable and certifiably safe controllers.

Finally, the real-world deployment of safe PIML still presents several challenges related to robustness, interpretability, and trustworthiness. While theoretical guarantees provide confidence in controlled settings, real-world systems exhibit uncertainties that are difficult to model precisely. Bridging the gap between theory and practice requires further research in hardware-in-the-loop safety validation, domain adaptation, and human-in-the-loop safety monitoring. Opportunities exist in creating benchmarks, standardized evaluation metrics, and open-source safety tools to facilitate broader adoption of safe learning-based control methods in real-world applications.

\section{Conclusions}

Safe learning for dynamical systems presents a promising pathway for enabling high-performance decision-making and control for a wide range of safety-critical applications. By integrating learning-based approaches with formal safety verification and robust control principles, it is possible to achieve adaptive and certifiably safe autonomous systems. However, several practical challenges remain, including the scalability of safety verification methods, reliable uncertainty quantification, and robust deployment in uncertain and faulty real-world scenarios. Addressing these challenges requires advancements in real-time safety filtering, risk-aware and constrained machine learning, and unified frameworks that systematically combines learning, verification, and control. The future of safe learning lies in developing scalable, interpretable, and provably safe methodologies that bridge the gap between theory and real-world deployment, ensuring reliable operation in safety-critical applications such as energy systems, critical infrastructure, robotics, autonomous driving, and industrial automation.

\bibliographystyle{IEEEtran}
\bibliography{references}

\addtolength{\textheight}{-12cm}   









\end{document}